\documentclass[aps,pre,twocolumn,groupedaddress,superscriptaddress,showpacs]{revtex4-1}

\usepackage{amsmath} \usepackage{amssymb} \usepackage{graphicx}
\usepackage{float} \usepackage{color}

\begin{document}

\title{Kinetic interfaces of patchy particles}

  \author{N. A. M. Ara\'ujo} \email{nmaraujo@fc.ul.pt}
\affiliation{Departamento de F\'{\i}sica, Faculdade de Ci\^{e}ncias,
Universidade de Lisboa, P-1749-016 Lisboa, Portugal, and Centro de
F\'isica Te\'orica e Computacional, Universidade de Lisboa, Avenida
Professor Gama Pinto 2, P-1649-003 Lisboa, Portugal}
    
  \author{C. S. Dias} \email{csdias@fc.ul.pt} \affiliation{Departamento
de F\'{\i}sica, Faculdade de Ci\^{e}ncias, Universidade de Lisboa,
P-1749-016 Lisboa, Portugal, and Centro de F\'isica Te\'orica e
Computacional, Universidade de Lisboa, Avenida Professor Gama Pinto 2,
P-1649-003 Lisboa, Portugal}

  \author{M. M. Telo da Gama} \email{mmgama@fc.ul.pt}
\affiliation{Departamento de F\'{\i}sica, Faculdade de Ci\^{e}ncias,
Universidade de Lisboa, P-1749-016 Lisboa, Portugal, and Centro de
F\'isica Te\'orica e Computacional, Universidade de Lisboa, Avenida
Professor Gama Pinto 2, P-1649-003 Lisboa, Portugal}

\pacs{82.70.Db,05.70.Ln,05.70.Fh,68.08.De,68.15.+e}

\begin{abstract}
We study the irreversible adsorption of patchy particles on substrates in the
limit of advective mass transport. Recent numerical results show that the
interface roughening depends strongly on the particle attributes, such as,
patch-patch correlations, bond flexibility, and strength of the interactions,
uncovering new absorbing phase transitions. Here, we revisit these results and
discuss in detail the transitions. In particular, we present new evidence that the
tricritical point, observed in systems of particles with flexible patches, is in the 
tricritical directed percolation universality class.
A scaling analysis of the time evolution of the correlation length for the
aggregation of patchy particles with distinct bonding energies confirms that
the critical regime is in the Kardar-Parisi-Zhang with quenched disorder
universality class.
\end{abstract}

  \maketitle

\section{Introduction}
The nonequilibrium evolution of growing interfaces has attracted many
experimental and theoretical
studies~\cite{Halpin-Healy1995,Vergeles1995,Wakita1997,Kim2001,Alava2004,Hallatschek2007,Sakaguchi2010,Takeuchi2010,Huergo2010,Huergo2011,Takeuchi2011,Takeuchi2012,Takeuchi2013,Takeuchi2014,Santalla2014}.
One of the most popular theoretical approaches considers kinetic discrete models to
describe particle aggregation on substrates. Albeit simple, these models are expected
to contain the relevant physics~\cite{Barabasi1995,Meakin1998,Odor2004}. In particular, the ballistic deposition model
(BD)~\cite{Vold1959,Vold1963}, is considered the prototype for
irreversible aggregation on substrates. In BD, the growth is driven solely by
the sequential addition of particles to the aggregate, which stick to the first
particle they touch, without subsequent rearrangement. From the simple rules of
ballistic deposition a complex structure emerges with a nontrivial porous bulk
structure (see e.g., Refs.~\cite{Vold1959,Barabasi1995,Blum2004}) and a
kinetically rough interface in the Kardar-Parisi-Zhang universality
class~\cite{Kardar1986,Barabasi1995}.

Inspired by recent advances in the production of patchy particles we have
proposed a stochastic model to study their aggregation on
substrates~\cite{Dias2013}, which in the limit of advective mass transport is a
generalized version of BD. Patchy particles are colloids with functionalized
surfaces, with new features such as selective and directional
particle-particle interactions, control over the valence, and the possibility
of forming permanent electrical
dipoles~\cite{Pawar2010,Sacanna2011,Sciortino2011,Iwashita2013,Iwashita2014,Markova2014,Sokolowski2014,Pizio2014}.
Studies of the irreversible aggregation on substrates reveal a nontrivial
dependence of the bulk and surface properties on the mechanism of mass
transport~\cite{Dias2013b}, on the strength of the patch-patch
interactions~\cite{Dias2013a,Dias2014a}, and on the spatial-patch
distribution~\cite{Dias2014}.

Here, we focus on the scaling properties of the growing interface in the limit
of advective mass transport. In this limit, we have found new absorbing phase
transitions depending on the patch-spatial arrangement~\cite{Dias2014} and a
crossover in the universality class of the interface depending on the relative
strength of the patch-patch interactions~\cite{Dias2014a}. These findings have been
discussed previously in the context of functional colloids. Here we revisit these
transitions and investigate their scaling properties in the framework of
kinetic discrete models of interfacial growth.

The paper is organized in the following way. In Sec.~\ref{sec::model} we
describe the model and recall some definitions. The two absorbing transitions
are discussed in Secs.~\ref{sec::patchpatch}~and~\ref{sec::flexibility}. The
crossover of the universality class of the interface is discussed in
Sec.~\ref{sec::strength}. Some final remarks and future perspectives are provided in
Sec.~\ref{sec::final}.

\section{Model}\label{sec::model}
In the ballistic deposition model~\cite{Barabasi1995,Odor2004,Alves2014}
particles are sequentially released from a position above the interface, chosen
uniformly at random, and move vertically towards the substrate sticking
irreversibly to the first particle they touch. To account for the directionality
of the interactions, the excluded volume interaction between particles, and the
short-ranged patch-patch attraction, characteristic of patchy colloids, we
proposed a generalized version of this model in Refs.~\cite{Dias2013,Dias2013b},
which we describe below.

To access larger-system sizes, let us consider a two-dimensional system of
patchy particles (disks) of unit diameter $\sigma$ with an initially empty flat
(linear) substrate of length $L$. As in the ballistic deposition model, we
iteratively generate a horizontal position, chosen uniformly at random above the
interface, to release a particle and follow its ballistic trajectory
downwards until the particle collides either with the substrate or with another particle.
Collisions with the substrate always result on adsorption of the particle at
the collision point with a random orientation.

\begin{figure}
\includegraphics[width=\columnwidth]{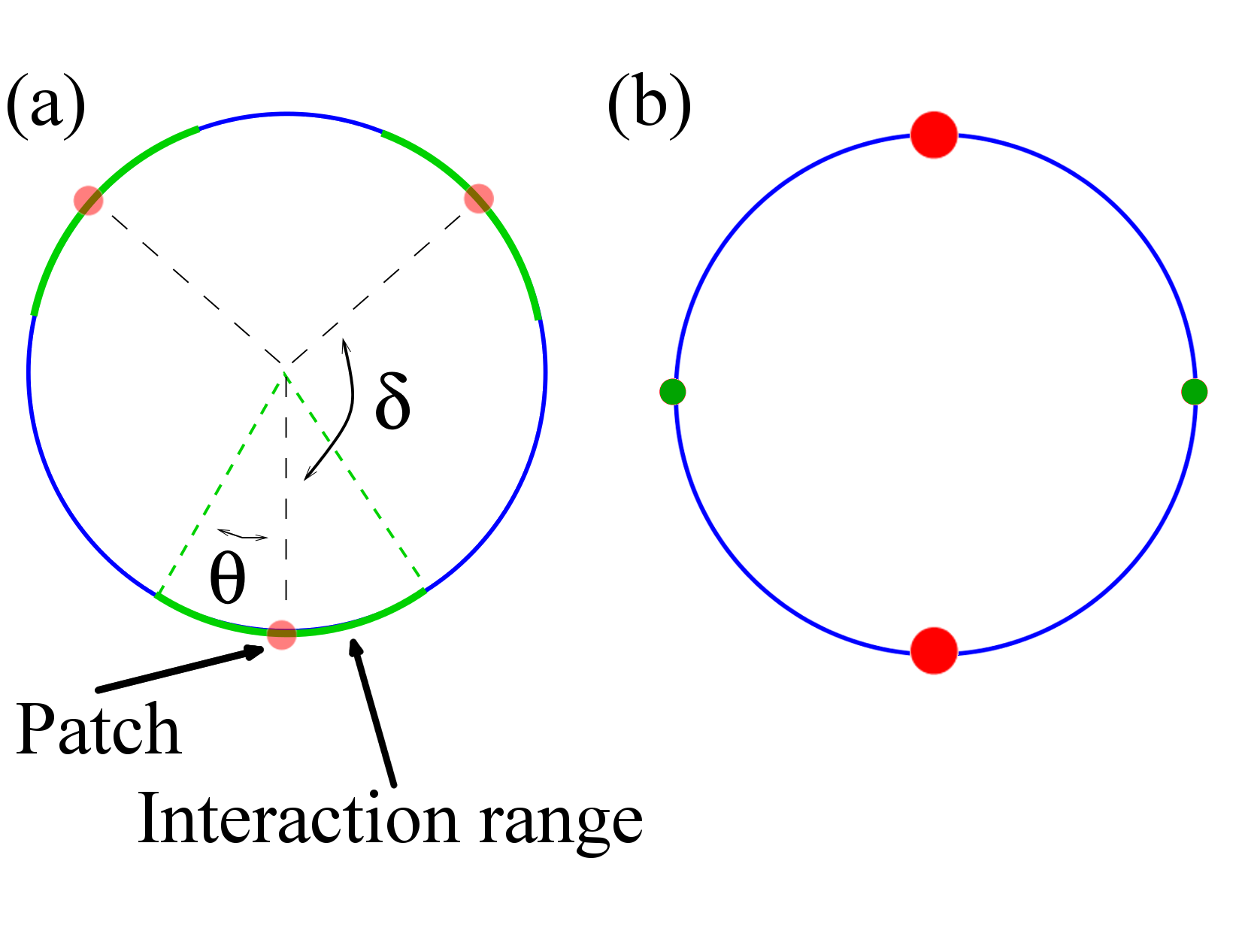}\\
\caption{(a) Patchy particle with three patches (red) on the surface and their
interaction range $\theta$ (green). The spatial arrangement of the patches is
described by an opening angle $\delta$, in units of $\pi\text{ rad}$, from the
center of the two adjustable patches to the center of the reference one. (b)
Four-patch particle with patches of two types: two $A$-patches on the poles and
two $B$-patches along the equator.  \label{fig.model}}
\end{figure}
While in the ballistic deposition model two particles stick together upon
collision, in the case of patchy particles the success of bond formation
depends on the relative orientation between the particles. The $n$ patches are
located on the surface of the particle and for each patch we define an
interaction range around it. The interaction range accounts for the extension
of the patch and the range of the patch-patch interaction and it is
characterized by a single parameter $\theta=\pi/6$, representing the maximum
angle with the center of the patch (see Fig.~\ref{fig.model}(a)). Two patches
bond in an irreversible way, a process we call binding, if their
interaction ranges partially overlap. Thus, stochastically, if the new particle
collides within the interaction range of a particle already on the substrate,
it binds to it with a probability $p=A_\mathrm{ir}/A$, where $A=\pi\sigma$ is
the surface area of the incoming particle and $A_\mathrm{ir}$ is the extension
of this area covered by the interaction range of all patches~\cite{Dias2013}.
If binding is not successful the particle is removed from the system and a new
one is released from the top.

\begin{figure}
\includegraphics[width=\columnwidth]{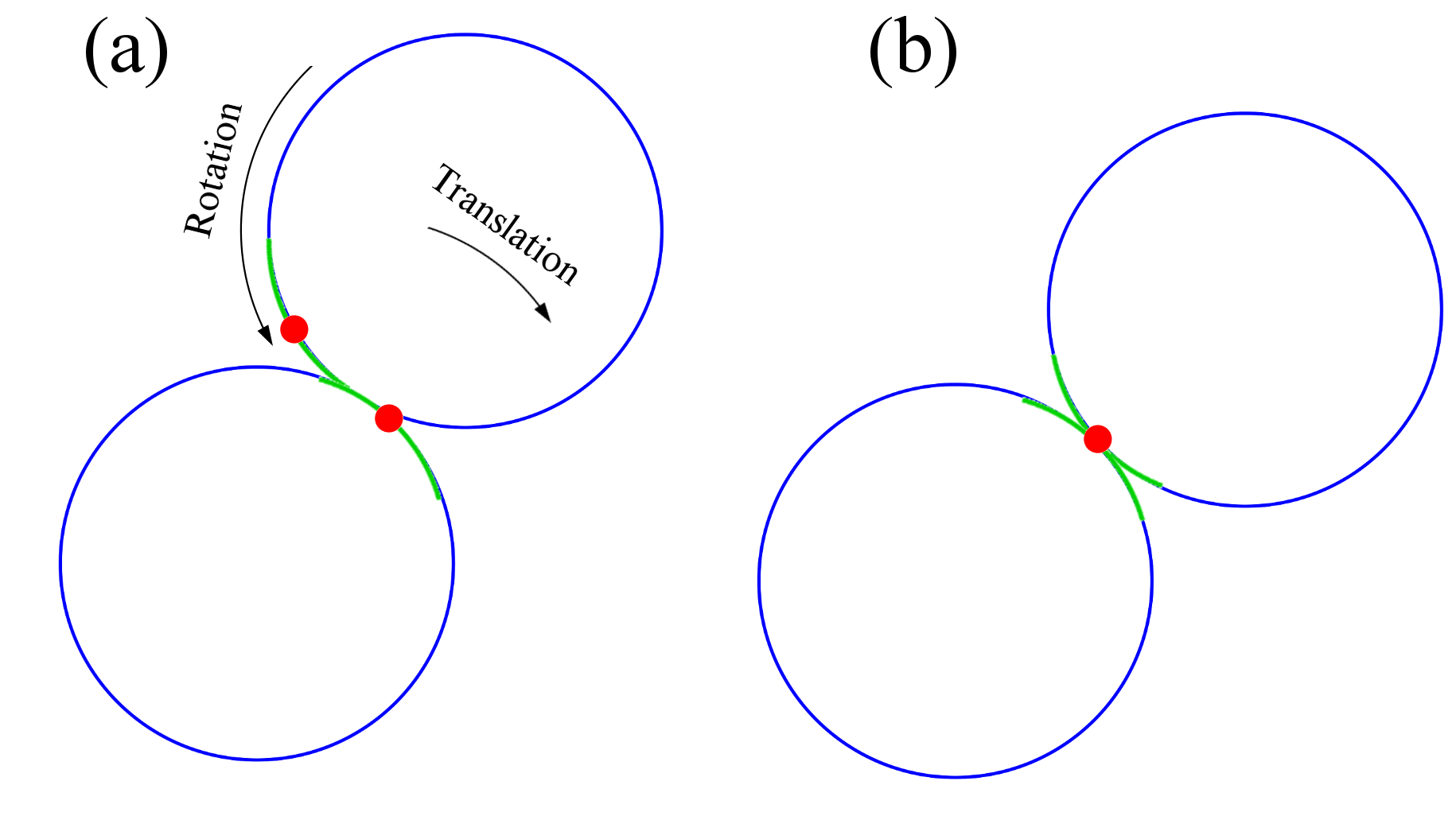}\\
\caption{(a) Successful binding between two patches occurs when their
interaction ranges partially overlap, in which case a bond is established
aligned along the two patches. Since the particle position and orientation in
the aggregate is considered irreversibly fixed, the alignment of the new
binding patches results solely from the rotation and translation of the
incoming particle, as shown in (b).  \label{fig.collision}}
\end{figure}
Inspired by chemical or DNA mediated
bonds~\cite{Geerts2010,Wang2012,Leunissen2011}, we consider highly directional
and very strong bonds between patches. Thus, in the case of successful binding,
the binding patch of the incoming particle is selected uniformly at random 
among its patches. The position and orientation of the incoming particle is
then adjusted such that the binding occurs along the center of the
two binding patches (see Fig.~\ref{fig.collision}).

\section{Patch-patch correlations}\label{sec::patchpatch}
The dependence of the interface of patchy particle aggregates on the spatial arrangement of the 
patches was studied in Ref.~\cite{Dias2014}. In that study we considered three-patch particles and 
investigated the dependence on the opening angle, showing that growth is suppressed below and above 
a minimum and a maximum opening angles, with two absorbing phase transitions between thick and
thin adsorbed film regimes. In this section, we investigate the nature of those transitions.

\begin{figure}
\includegraphics[width=\columnwidth]{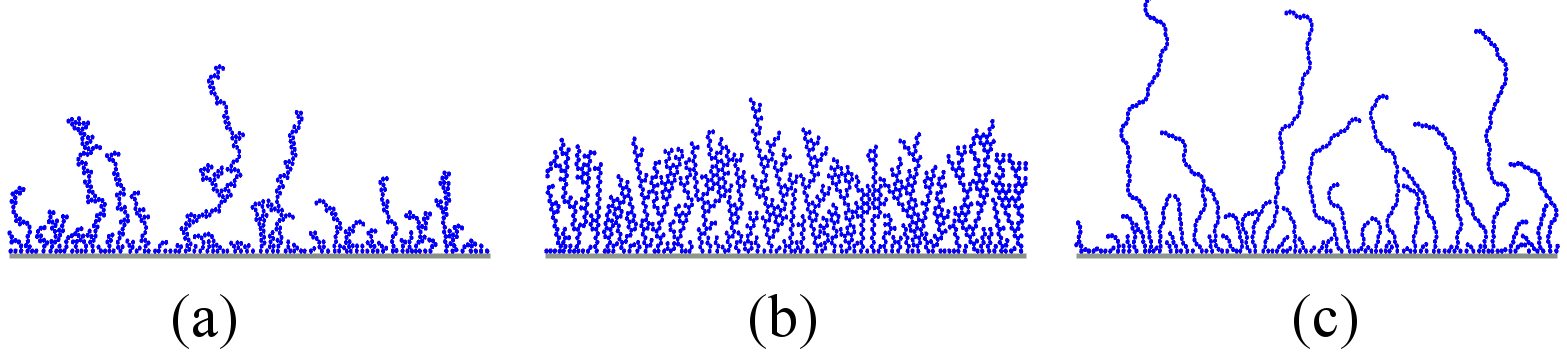}\\
\caption{Snapshot of networks of patchy particles for different values of
$\delta$: (a) $0.4\pi$, (b) $2\pi/3$, and (d) $0.85\pi$. \label{fig.snapshot}}
\end{figure}
As shown schematically in Fig.~\ref{fig.model}(a), the particles are disks with three
patches: one reference patch and two adjustable ones. The spatial arrangement
of the patches is characterized by the opening angle $\delta$, from the reference
patch to the adjustable ones. For simplicity, we define the units of $\delta$
as $\pi\text{ rad}$. As the particles are sequentially added, a network of patchy
particles grows away from the substrate (see Fig.~\ref{fig.snapshot}) but, its
growth is only sustained for $\delta_\mathrm{min}<\delta<\delta_\mathrm{max}$. 

\begin{figure}
\includegraphics[width=\columnwidth]{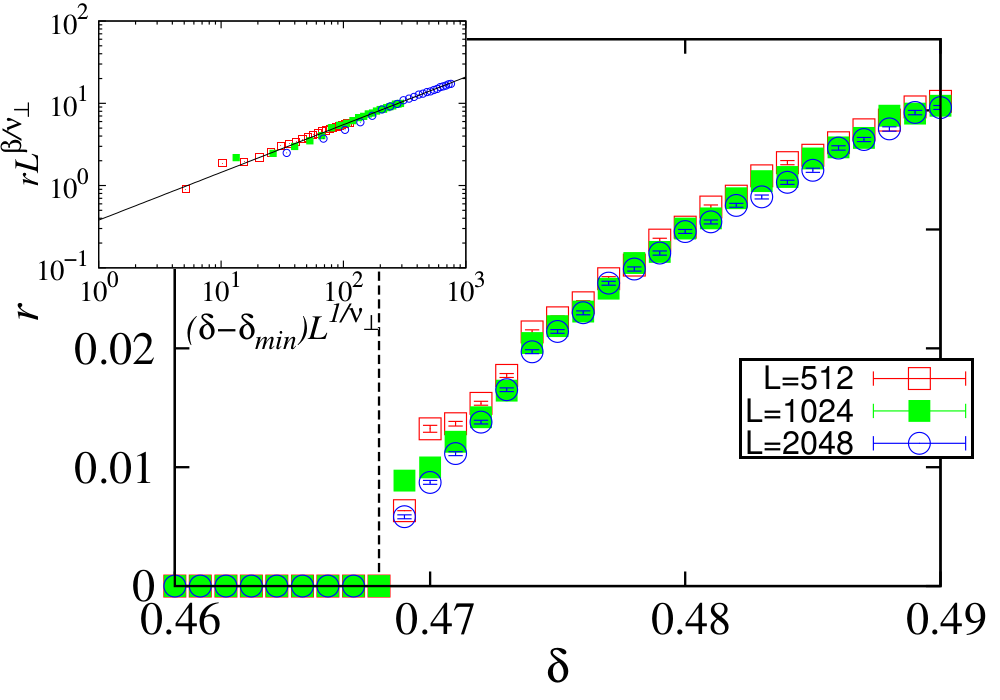}\\
\caption{Fraction $r$ of successful binding attempts in the stationary state as
a function of $\delta$, for three different substrate lengths $L$, namely,
$512$, $1024$, and $2048$. The inset shows the finite-size scaling in
logarithmic scale using the exponents of the directed percolation universality
class. We considered $\beta_\mathrm{DP}=0.58$ and $\nu_\mathrm{DP}=0.73$.
Results are averages over $\{400,200,100\}$ samples for $L=\{512,1024,2048\}$.
\label{fig.first.transition}}
\end{figure}

For $\delta<\delta_\mathrm{min}$, the angle between the patches is such that all
patches are in the same hemisphere. Thus, the patches of particles in the aggregate
are pointing most likely towards the substrate and are not accessible to new
incoming particles. When there are no more patches available to establish
new bonds, the growth is suppressed. A systematic finite-size study for the
threshold value gives $\delta_\mathrm{min}=0.468\pm0.001$~\cite{Dias2014}. This
threshold is above that expected from purely geometrical arguments,
revealing strong collective effects~\cite{Dias2014}. To analyze the transition
to the absorbing state at $\delta_\mathrm{min}$, we define as the order parameter
$r$ the fraction of successful binding attempts.
Figure~\ref{fig.first.transition}, shows the dependence of $r$ on $\delta$ for
different system sizes. Clearly, in the thermodynamic limit, $r$ vanishes at
$\delta_\mathrm{min}$ and it grows continuously with $\delta$. As shown in the
inset, a data collapse is obtained for the finite-size scaling consistent with
the directed percolation (DP) universality class in two
dimensions~\cite{Henkel2008b,Lubeck2003,Odor2004}. 

\begin{figure}
\includegraphics[width=\columnwidth]{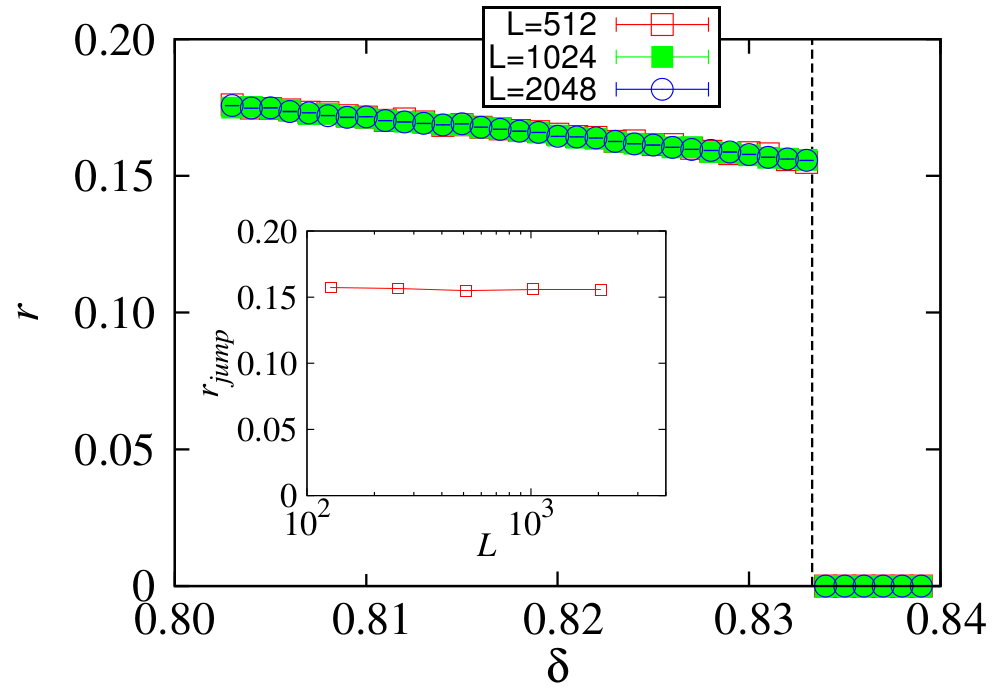}
\caption{Fraction $r$ of successful binding attempts in the stationary state as
a function of $\delta$, for three different substrate lengths $L$, namely,
$512$, $1024$, and $2048$. The inset shows the dependence of the size of the
jump on the system size $L$. Results are averages over
$\{1600,800,400,200,100\}$ samples for $L=\{128,256,512,1024,2048\}$.
\label{fig.second.transition}}
\end{figure}

For $\delta>\delta_\mathrm{max}$, the distance between the adjustable patches
is such that, if one particle binds to one of these patches it shields the access 
of a new particle to the other. Consequently, branching is suppressed and only
chains grow away from the substrate. These chains are locally tilted and their
growth direction (given by the available patches), fluctuates while growing.  
Eventually, the growing tip of the chain points down and its growth
is suppressed. The absorbing state occurs when all tips are either pointing
down or covered by other chains. From geometrical arguments one expects
$\delta_\mathrm{max}=5/6$, a value that was numerically
confirmed~\cite{Dias2014}.  Figure~\ref{fig.second.transition} shows the
dependence of $r$ on $\delta$ close to this second transition. By contrast to
the transition at $\delta_\mathrm{min}$, at $\delta_\mathrm{max}$ the
transition is discontinuous and the growth rate jumps at the threshold. In the
inset of Fig.~\ref{fig.second.transition} it is clear that the jump does not
vanish in the thermodynamic limit, discarding the existence of strong
finite-size effects.

For $\delta_\mathrm{min}<\delta<\delta_\mathrm{max}$ a ramified network of
patchy particles grows from the substrate in a sustained way. In the stationary
state, the interface is always in the Kardar-Parisi-Zhang universality
class~\cite{Kardar1986,Dias2014} as was observed for isotropic
sticking particles~\cite{Jullien1989,Meakin1998}. However, for patchy
particles the saturation roughness shows a non-monotonic dependence on
$\delta$, with a minimum at $\delta=3/2$. This minimum occurs when the three
patches are equidistant, which favors branching and consequently leads to a
decrease of the roughness.

\section{Bond flexibility}\label{sec::flexibility}

\begin{figure}
\begin{center}
\includegraphics[width=0.6\columnwidth]{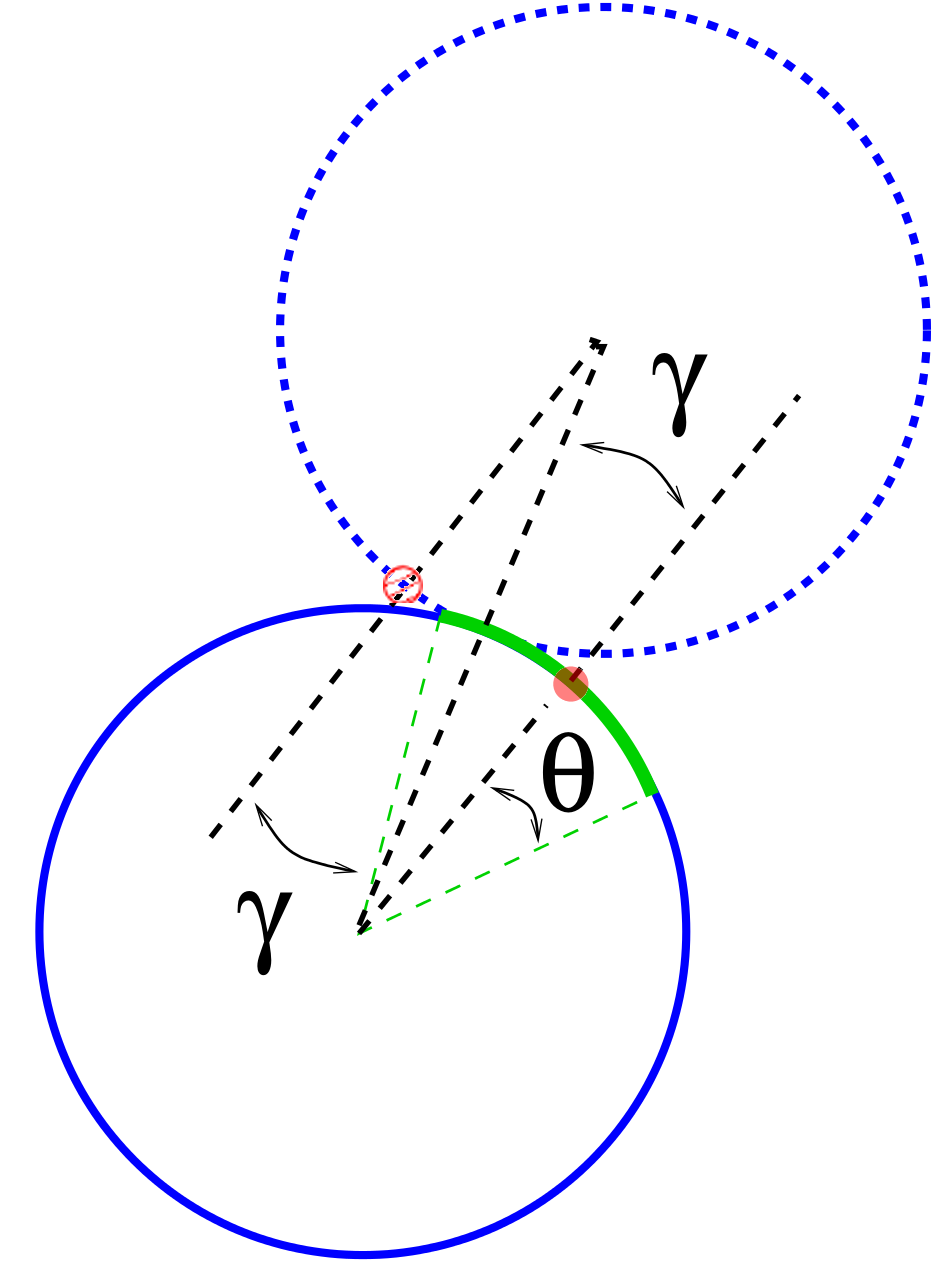}
\end{center}
 \caption{Non-optimal binding between two patchy particles where the bond
orientation deviates from the optimal orientation by an angle $\gamma$.  For
simplicity, we consider the same deviation for both patches and that the sense
of rotation is always from the center of the patch to the point of
collision.~\label{fig.flexibility}} 
\end{figure}
For simplicity, in the previous section the position of the incoming particle
is adjusted such that the center of the colloids and of their patches is
perfectly aligned. However, in reality, one expects some degree of flexibility
around this optimal orientation~\cite{Loweth1999,Geerts2010}. A simple strategy
to account for flexibility was proposed in Ref.~\cite{Dias2014}, which
takes advantage of the stochastic nature of our model. The idea is still to
consider rigid and irreversible bonds but, at a binding event, the orientation
of the bond deviates by an angle $\gamma$ from the optimal orientation (see
Fig.~\ref{fig.flexibility}). The value of $\gamma$ is drawn randomly from a
Gaussian distribution of zero mean and dispersion $F\theta$, where $F$ is the
flexibility, truncated at $F\theta$. The sense of rotation of $\gamma$ is
always from the center of the patch to the point of collision.

\begin{figure}
\includegraphics[width=\columnwidth]{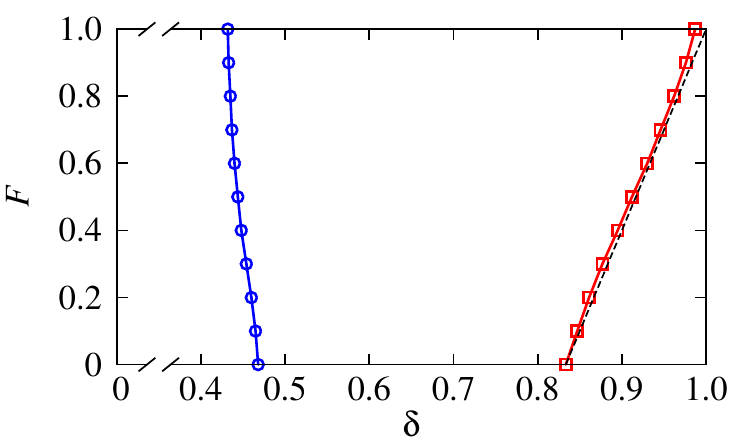}
\caption{Phase diagram in the space of flexibility ($F$) and opening angle
($\delta$). The solid lines correspond to the lower ($\delta_\mathrm{min}$) and
upper ($\delta_\mathrm{max}$) thresholds.  The data points are extrapolations
to the thermodynamic limit from the size dependence of the thresholds. The
dashed line is the theoretical prediction for $\delta_\mathrm{max}$ based on
geometrical arguments.\label{fig.diagram.flexibility}}
\end{figure}
\begin{figure}
\includegraphics[width=\columnwidth]{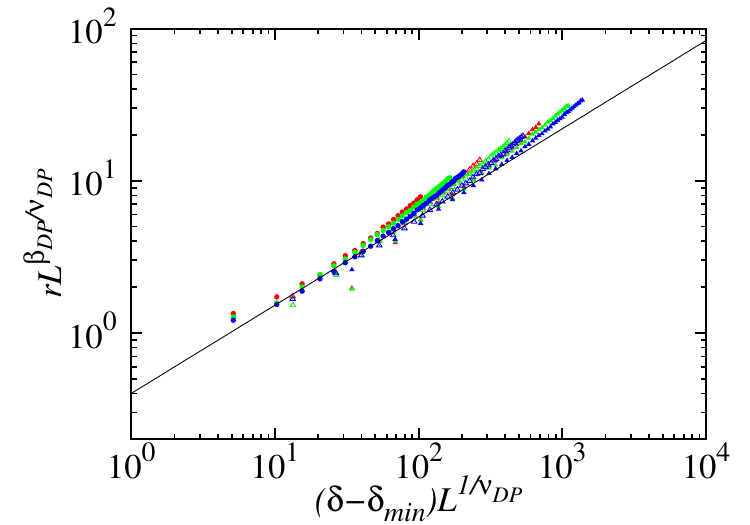}
\caption{Finite-size scaling of the order parameter $r$ for different values
of the flexibility, $F=\{0.2,0.4,0.6\}$, system size, $L=\{512,1024,2048\}$, averaged 
over $\{400,200,100\}$ samples. We considered
$\beta_\mathrm{DP}=0.58$ and $\nu_\mathrm{DP}=0.73$ consistent with
the DP universality class in two dimensions.\label{fig.lower.transition}}
\end{figure}
Results for different values of the flexibility are summarized in the diagram shown
in Fig.~\ref{fig.diagram.flexibility}. The active region does widen with the
flexibility but there is always a lower and an upper thresholds. For the range
of flexibilities considered, the transition at $\delta_\mathrm{min}$ is always
continuous and in the DP universality class, as shown in
Fig.~\ref{fig.lower.transition}. The threshold $\delta_\mathrm{min}$ decreases
with $F$ but it is always significantly higher than $1/3$, the one predicted
from purely geometrical considerations.

\begin{figure}
\includegraphics[width=\columnwidth]{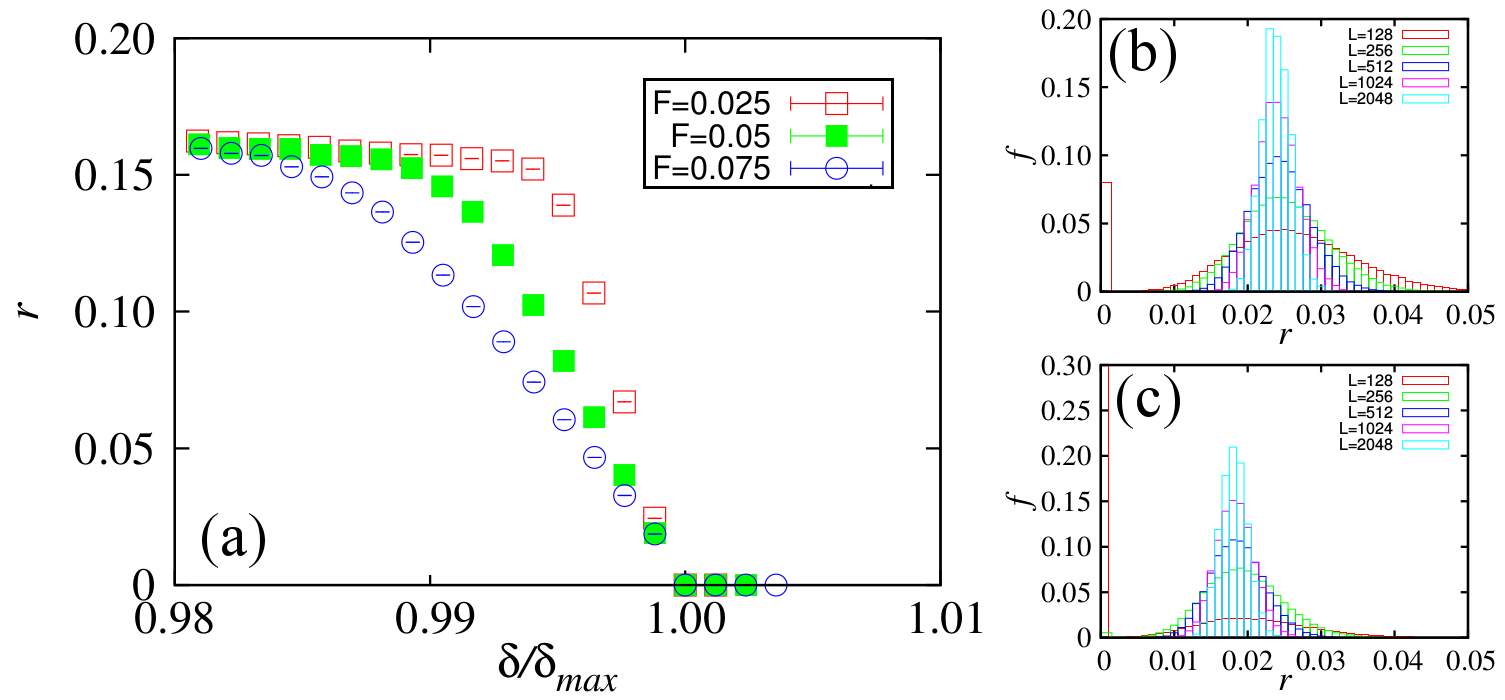}
\caption{(a) Order parameter $r$ as a function of $\delta/\delta_\mathrm{max}$,
for $F=\{0.025,0.05,0.075\}$, $L=2048$ and averaged over $100$ samples. (b)
and (c) are the histograms of the order parameter at the threshold
$\delta_\mathrm{max}$ for $F=0.025$ and $F=0.075$, respectively, showing that
the jump does not vanish with the system size. Results are averaged over
$\{160000,80000,40000,20000,10000\}$ samples for $L=\{128,256,512,1024,2048\}$.
\label{fig.higher.trans}}
\end{figure}
\begin{figure}
\includegraphics[width=\columnwidth]{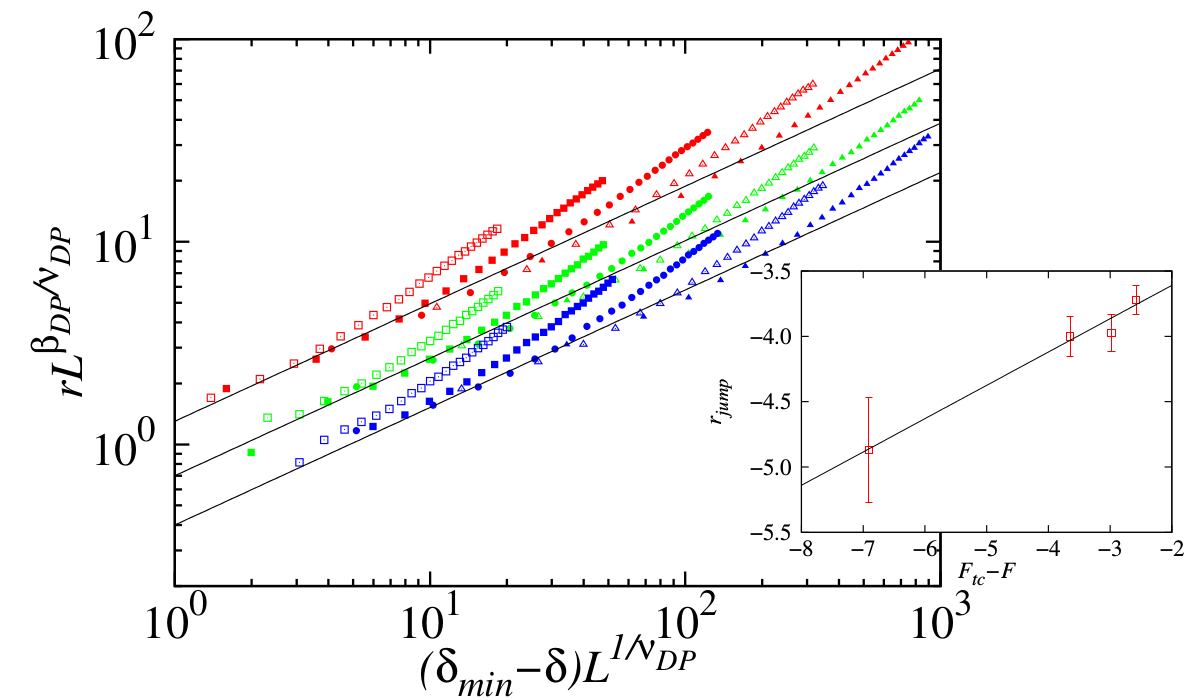}
\caption{Finite-size scaling of the order parameter $r$ for
$F=\{0.2,0.4,0.6\}$, and three different substrate lengths
$L=\{512,1024,2048\}$, averaged over $\{400,200,100\}$ samples. The inset
shows the dependence of the size of the jump $r_\mathrm{jump}$ on the distance to the
tricritical flexibility $F_\mathrm{tc}$, for $F<0.1$ and $F_\mathrm{tc}=0.1$.
The solid line is a guide to the eye scaling as
$(F_\mathrm{tc}-F)^b$, with $b=0.25$.~\label{fig.higher.trans.cont}}
\end{figure}
For the second transition, the threshold $\delta_\mathrm{max}$ increases with
$F$ and converges towards unity at large flexibilities. In fact, the threshold
value is well approximated by,
\begin{equation}
\delta_\mathrm{max}(F)= \delta_\mathrm{max}(0)+\frac{F\theta}{\pi} \ \ ,
\end{equation}
obtained from geometrical arguments~\cite{Dias2014}, corresponding to the
dashed line in Fig.~\ref{fig.diagram.flexibility}. The nature of the transition
also changes with $F$. Figure~\ref{fig.higher.trans}(a) shows the dependence of
the order parameter $r$ on $\delta$ close to the second transition, for three
different values of $F$.  The numerical results suggest that even for
$F=0.075$, the transition is still discontinuous (see also the histogram of the
order parameter in Fig.~\ref{fig.higher.trans}(b)~and~(c)) but the size of the
jump decreases as $F$ increases. For $F>0.2$ the numerical results clearly suggest 
a continuous transition in the DP universality class (see
Fig.~\ref{fig.higher.trans.cont}). The inset of
Fig.~\ref{fig.higher.trans.cont} suggests that the size of the jump
decreases with the distance to $F=0.1$ as a power law with exponent $0.25$.
This suggests, in turn, that the nature of the transition changes from discontinuous to
continuous at a tricritical flexibility $F_\mathrm{tc}\approx0.1$, in the
tricritical directed percolation universality
class~\cite{Lubeck2006,Grassberger2006}.

\section{Weak and strong bonds}\label{sec::strength}
Patchy particles with distinct patch-patch interactions yield interesting
bulk properties at equilibrium~\cite{Tavares2009,Tavares2009a,Tavares2010a,Tavares2010,Russo2011,Russo2011a,Almarza2011,Almarza2012,DeLasHeras2012,Tavares2014}
and novel nonequilibrium interfaces~\cite{Dias2013a,Dias2014a}. The typical strategy is
to consider $2AnB$ particles, with two strong $A$- and $n$ weak $B$-patches. The
$A$-patches are in the poles, while the $B$-patches are equally spaced along the
equator (see Fig.\ref{fig.model}~(b)). A generalization of our model to account
for these two energy scales was proposed in Refs.~\cite{Dias2013a,Dias2014a},
based on dissimilar binding probabilities that can be formally related to the
activation energy of binding. At a collision event with partial overlap between
the interaction range of two patches $i$ and $j$, the binding is successful
with a binding probability $P_\mathrm{ij}$ that depends on the type of patch
pair $ij$. In particular, since $A$-patches are stronger than $B$-patches, we
consider $P_\mathrm{AA}>P_\mathrm{BB}=P_\mathrm{AB}$. Without loss of
generality, we consider $P_\mathrm{AA}=1$ and define the sticking coefficient
$r_\mathrm{AB}=P_\mathrm{AB}/P_\mathrm{AA}$. Lower sticking coefficients
favor chain growth over branching.

In Ref.~\cite{Dias2014a} we have shown that a crossover
of the interfacial roughening from the Kardar-Parisi-Zhang (KPZ) to
the KPZ with quenched disorder (KPZQ) is observed when $r_\mathrm{AB}$ is sufficiently
small, i.e., when the $A$-patches are significantly stronger than the
$B$-patches. For $r_\mathrm{AB}\ll1$, the strong $A$-patches, promote growth along the poles, 
favoring the aggregation of $AA$ chains. These
chains are likely to extend over long lateral regions. These long chains expose
their $B$-patches and shield the access to any underlying $A$-patch, thus the
probability of binding there is significantly lower. This is expected to have a similar effect
to that of quenched noise~\cite{Dias2014a}.

Note that the KPZ universality class is very robust~\cite{Kardar1986}, while
KPZQ is only observed at the critical depinning
transition~\cite{Olami1994,Csahok1993,Leschhorn1997,Amaral1995,Amaral1994,Buldyrev1992,Halpin-Healy1995,Tang1992}.
Nevertheless, for patchy particles with weak and strong bonds one
finds remarkably an entire critical region of $r_\mathrm{AB}$ where KPZQ is observed. This
is likely due to the balance of two competing mechanisms that keep the
system at criticality. As $r_\mathrm{AB}$ decreases, the probability of
binding to $B$-patches decreases and the growth of longer $AA$-chains increases.
The longer these chains, however, the larger is the number of $B$-patches available for bonding, which compensates 
the decrease in the binding probability.

Recent experiments on aggregation of ellipsoids at the edge of an
evaporating drop suggest that, for sufficiently large major-minor axis
aspect ratio the interface is also in the KPZQ universality
class~\cite{Yunker2013}.  Together with these experiments, a discrete
model was proposed to argue that KPZQ was driven by a colloidal Matthew
effect~\cite{Yunker2013}. However, numerical simulations of the same
model by Nicoli \textit{et al.}~\cite{Nicoli2013} suggest a different
interpretation of the theoretical results. Oliveira and Reis performed a
careful statistical analysis of the correlation length and concluded
that the differences observed in the growth and roughness exponents are
due, instead, to a crossover to columnar growth~\cite{Oliveira2014}.
Although the results for the interface of the simple models appear to be
settled the nature of the interface of the experimental system is still
an open problem~\cite{Yunker2013}.

\begin{figure}
\includegraphics[width=\columnwidth]{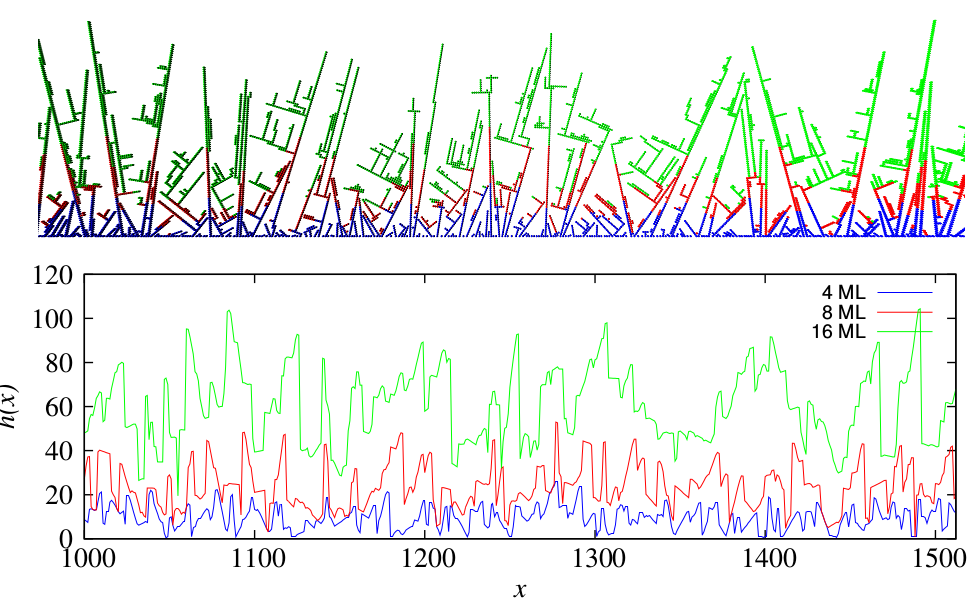}
\caption{Snapshot of a section of the aggregate (top) and height profile
(bottom) after the adsorption of four (blue), eight (red), and $16$
(green) monolayers of $2A2B$ particles, with $r_\mathrm{AB}=0.01$.
Results are for one single sample and a substrate of length
$L=4096$.~\label{fig.height.profile}}
\end{figure}
\begin{figure}
\includegraphics[width=\columnwidth]{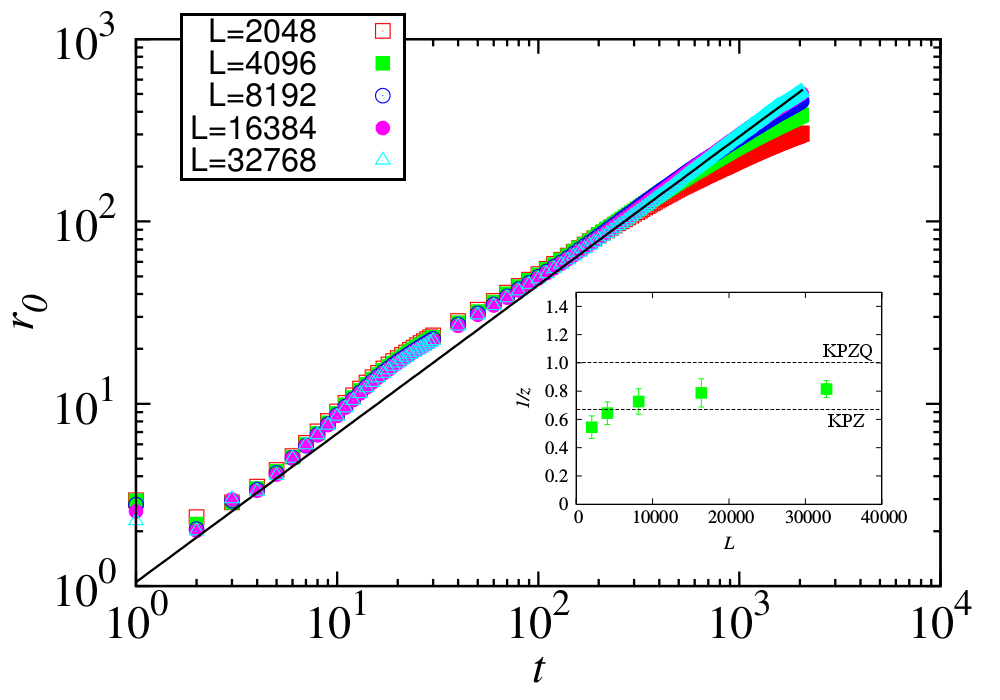}
\caption{Scaling of the first zero of the autocorrelation function with
time for four different substrate lengths,
$L=\{2048,4096,8182,16384,32768\}$, averaged over
$\{20000,20000,10000,10000,2000\}$ samples. The inset depicts the size
dependence of the slope $1/z$, where error bars are given by the
standard deviation of the local slope in the interval considered for the
fitting. The dashed lines correspond to the expected value for KPZ
(bottom) and KPZQ (top) universality
classes.~\label{fig.correlation.length}}
\end{figure}

In order to proceed and to discard the possibility of columnar growth in
the model of patchy particles with strong and weak bonds, we performed
the analysis proposed in Ref.~\cite{Oliveira2014}.
Figure~\ref{fig.height.profile} shows a section of the aggregate and
height profile for one sample with $r_\mathrm{AB}=0.01$. The time
evolution of the height profile reveals no evidence of columnar growth.
A more quantitative analysis is illustrated in
Fig.~\ref{fig.correlation.length} where we have plotted the scaling of
the position $r_0$ of the first zero of the autocorrelation function
with time. This position is expected to scale as,
\begin{equation}
r_0\sim t^{1/z} \ \ ,
\end{equation}
where $z$ is the dynamic exponent. In spite of the strong finite-size
effects, the size scaling of the slope (inset of the figure) indicates
that the growth is consistent with the KPZQ universality class.

\section{Final remarks}\label{sec::final}
We reviewed our recent results on the kinetic roughening of interfaces of patchy
particles and showed that the scaling of the interface depends strongly on the
patch-patch correlations, the bond flexibility, and the strength of interactions. For
particles with patch-patch correlations and bond flexibility we found two absorbing phase
transitions that are, in general, of different nature. While the first transition is always
continuous in the directed percolation universality class the second is
either continuous or discontinuous depending on the bond flexibility. A scaling
analysis of the size of the jump close to the tricritical flexibility suggests
that the tricriticality is in the tricritical directed percolation
universality class. For particles with weak and strong bonds we analyzed the kinetics of
particles with two types of patches ($A$ and $B$). When the strength of the $A$ and $B$ patches is similar, the interface roughness is in the Kardar-Parasi-Zhang
universality class. However, for $A$-patches significantly stronger than 
the $B$-patches, the interface is in the universality class of
Kardar-Parasi-Zhang with quenched disorder. This critical universality class is
observed for an entire range of the relative strength of the interactions.

Beyond their theoretical interest, our findings have three consequences with practical applications. 
First, we have shown that the roughness of the interface can be controlled by the spatial distribution 
of patches or the relative strength of their interactions. Second, we revealed that sustained growth 
is only possible for certain patch arrangements. Third, the existence of an
extended region of the parameter space where the critical KPZQ universality
class is observed opens the possibility for an experimental realization of such systems using patchy colloids with
weak and strong bonds.

\begin{acknowledgments} 
We acknowledge fruitful discussions with A. G. Yodh and T. J. Oliveira, as well
as financial support from the Portuguese Foundation for Science and Technology
(FCT) under Contracts nos. EXCL/FIS-NAN/0083/2012, PEst-OE/FIS/UI0618/2014, and
IF/00255/2013. 
\end{acknowledgments}

\bibliography{colloids}

\end{document}